\documentclass[12pt,preprint]{aastex}

\newcommand{\jcd}{Christensen-Dalsgaard}

\shorttitle{Helioseismology and solar abundances}
\shortauthors{Antia and Basu}

\begin{document}

\title{The discrepancy between solar abundances and helioseismology}

\author{H. M. Antia}
\affil{Tata Institute of Fundamental Research, Homi Bhabha Road,
Mumbai 400005, India}
\email{antia@tifr.res.in}

\and

\author{Sarbani Basu}
\affil{Astronomy Department, Yale University, P. O. Box 208101,
New Haven CT 06520-8101, U.S.A.}
\email{basu@astro.yale.edu}

\begin{abstract}
There have been recent downward revisions of the solar photospheric
abundances of Oxygen and other heavy elements.
These revised abundances along with
OPAL opacities are not consistent with seismic constraints.
In this work we show that
the recently released OP opacity tables cannot resolve this
discrepancy either. While the revision in opacities does not seem to
resolve this conflict, an upward revision of Neon abundance
in solar photosphere  offers a possible solution to this problem.

\end{abstract}

\keywords{Sun: abundances --- Sun: oscillations --- Sun: interior}

\section{Introduction}

Recent analyses of spectroscopic data using modern three dimensional
hydrodynamic atmospheric models 
have suggested that the solar abundance of Oxygen and other abundant elements
needs to be revised downward (Allende Prieto, Lambert \& Asplund 2001, 2002; 
Asplund et al.~2004a,b; Melendez 2004). Asplund et al.~(2004a) claim that the 
Oxygen abundance should be reduced by a factor of about 1.48 from the earlier estimates of
Grevesse \& Sauval (1998). The abundances of C, N, Ne, Ar and other
elements are also reduced (Asplund et al.~2004b).
As a result, the ratio (by mass) of heavy element to 
hydrogen abundance, $Z/X$, reduces from 0.023 to 0.0165, which causes
the heavy element abundance
in the solar envelope to reduce from $Z=0.017$ to 0.0122.
This will cause the opacity
of solar material to decrease, in turn reducing the depth of the
convection zone (henceforth CZ) in solar models.
Bahcall \& Pinsonneault (2004) have
constructed a standard solar model using these revised abundances to find that
the depth of the CZ is indeed  reduced significantly making it
inconsistent with seismically established value. Basu \& Antia
(2004) and Bahcall et al.~(2004a, 2005) also attempted to study
solar models with reduced abundances and found that opacities
near the base of the convection zone needs to be increased by
10--20\% to make them consistent with seismic data.
Turck-Chi\`eze et al.~(2004) also find that solar models constructed
using revised heavy element abundances are not consistent with
seismic inferences.
Subsequently, Asplund et al.~(2004b) have revised the
abundances of some elements still further. As a result, the
discrepancy can be expected to be somewhat larger.

Since the discrepancy can be attributed to reduced opacities, there
is a need to examine the opacity calculations. Recently,
Seaton \& Badnell (2004) and Badnell et al.~(2004) have carried out independent opacity
calculations under the OP project. 
Near the base of the solar CZ, they only find a  2\% increase
in opacity as compared to OPAL.
This is not likely to resolve the discrepancy, but nevertheless,
the effect needs to be studied in detail.
Since two completely independent opacity calculations agree very
well with each other, it is unlikely that any possible revision
in opacity will be large enough to address the discrepancy caused
by the downward revision of solar abundances.
Nevertheless, additional independent tests of opacity calculations are desirable
to ensure reliability of opacities which are crucial input to stellar
model calculations.

In this work we also study the effect of varying abundances of  many heavy elements
separately to check which of them are effective
in addressing the discrepancy.
Such studies will help in identifying the elements which play a crucial
role in the discrepancy in solar models.
It turns out that the
problem may be resolved if the Neon abundance is increased
by about 0.6 dex, i.e., by a factor of 4. It may be noted that
the Neon abundance in the photosphere cannot be determined
spectroscopically and hence the uncertainties could be large.

\section{The technique}

Following Basu \& Antia (2004) we construct solar envelope models
with different heavy element abundances. All these models
have the seismically estimated hydrogen abundance, $X$, of 0.739
(Basu \& Antia 1995, 2004) and the depth of the convection
zone of $0.2867R_\odot$ (\jcd\ et al.~1991; Basu \& Antia 1997, 2004).
As a result there are no free parameters in these models and the
sound speed and density in these models can be compared with
seismically inferred values to check for consistency.
These models use the OPAL equation of state (EOS) (Rogers \& Nayfonov 2002)
and opacity tables from OPAL (Iglesias \& Rogers 1996) or OP
(Badnell et al.~2004). In all cases, the opacity has been
calculated using the appropriate mixture of heavy elements, while
the EOS has been calculated using a standard mixture for which the
OPAL tables are available.
In principle, the EOS tables also need to be modified
in view of the change in mixture of heavy elements. That has not been done
since the EOS is
not particularly sensitive to the detailed breakup of heavy element abundance.

We construct solar envelope models using the heavy element abundances
as given by Grevesse \& Sauval (1998) (referred to as GS98) or
Asplund et al.~(2004b) (referred to as Asp04). Furthermore, to study
the effect of abundances of individual heavy elements, we have
constructed models where the abundances of C, N, O, Ne, Mg, Si, S
and Fe are separately reduced as compared to their abundances
in GS98. For each solar envelope model we compare the density
profile in the lower part of the convection zone with that inferred through
seismic inversions. In general, the two do not agree with each
other and we need to modify the opacity near the base of the
convection zone to get agreement. The opacity modification
required to get the density to agree 
is a measure of consistency between seismic data, abundances
and opacities.

\section{Results}

We compare the density profile in each solar envelope model with
seismically inverted density and the results for some of the models
are shown in Fig.~1. It is clear that the density profile in  models
constructed using the revised abundances is significantly different,
from that of the Sun, the densities being higher than solar density.
The estimated errors in the  density inversion results in
the lower CZ is about 1.5\%, including systematic errors due to
uncertainties in X, EOS etc. (Basu \& Antia 2004).
The difference in density is more than 15\%, much larger than the errors.
On the other hand, models with GS98 abundances have density profiles within error
limits, thus these  models are consistent with seismic constraints.
For the Asp04 mixture the opacities need to be increased by
$(27\pm3)$\% for OPAL models and $(25\pm3)$\% for OP models to get
the density profile in agreement with seismic inversions.
The OP and OPAL models for the same mixture are close to each other
and the difference between them are consistent with
known differences in opacity estimates.

It is clear from Fig.~1 that there
is not much difference between models using OPAL and OP opacities.
In order to quantify the differences in opacities, Fig.~2 shows
the relative difference between the two opacities as a function of temperature
in a solar model using Asp04 mixture. The differences are taken
at the same density and temperature and reflect the differences in
the actual opacity calculations. The differences are generally
less than 2\% in the radiative interior.
Near the base of the CZ the difference is of order of 1\%.

Since the two independent opacity calculations agree remarkably
well with each other, it is unlikely that the discrepancy caused
by reduced abundances can be due to uncertainties in opacities.
To separate out the contribution of each element we construct
a series of solar envelope models with the GS98 mixture, with the abundance
of one element at a time reduced by the amount shown in the
third column of Table~1.
Two of these models (for reduction in O and Fe abundances)
are also shown in Fig.~1.
The results using OPAL opacity tables are summarized in Table 1,
which lists
the required opacity modification to restore the density profile
to that inferred by inversions. It also lists the logarithmic
derivative of the required opacity modification with respect to
abundance of each element. It is clear that the derivative
is significant for abundant elements like O, Ne, Fe.
Thus the required opacity modification can be controlled by
adjusting the abundance of some of these elements. And we see
from Fig.~1, that the model with the increased Ne abundance
is actually consistent with seismic results.
If we believe that 
the abundance determination of O, Fe have improved significantly
in recent times, then there may not be much uncertainty in their
estimated abundance and we do not have much freedom to vary those
abundances.  The photospheric abundance 
of Ne  however, may involve higher uncertainties since it
can not be determined spectroscopically due to lack of
suitable photospheric lines and has to be  determined
from coronal lines. This could also involve uncertainties
due to possible  fractionation in these layers, as
the coronal  Ne abundance may not reflect that in the photosphere.
Thus we speculate that the effect of reduction in abundance
of other elements like C, N, and O may be compensated by an
increase in the Ne abundance. The required Ne abundance can be
estimated by constructing models with different values of Ne
abundance to estimate the required abundance to match the density
profile. It turns out that we need an increase in Ne abundance by
$0.63\pm0.06$ dex when OP opacities are used, and by $0.67\pm0.06$ dex when OPAL
opacities are used. This corresponds to an increase in abundance by
a factor of just over 4. 
It can be seen from Fig.~1 that the envelope models constructed using these
abundances have the correct density profile.
We can also estimate the required increase in Ne abundance from the
partial derivative given in Table~1, but that gives a somewhat larger
estimate, presumably because the derivative itself would increase when
Ne abundance increases by a factor of 4.

Primary seismic inversions for sound speed and density are
independent of opacities, but we need to use opacities in order to infer the temperature
and hydrogen abundance profiles in the solar interior
(Gough \& Kosovichev 1988; Shibahashi \& Takata 1996; Antia \& Chitre 1998).
We therefore check the differences in the inferred
temperature and $X$ profile of the Sun arising from  the use of the 
two different opacity tables. Figure~3 shows the difference
in temperature and hydrogen abundance inferred using the
technique described by Antia \& Chitre (1998). The differences
are taken between the profiles calculated using OP and OPAL
opacities (in the sense OP $-$ OPAL) when the same
heavy elements abundances are used. The differences of
less than 0.005 are comparable to estimated errors from other
sources. In contrast, the difference between the $T$ and $X$ profiles
obtained using GS98 and
Asp04 mixtures with the same opacity tables are an order of magnitude larger.
Thus the differences in opacities between OPAL and OP
do not lead to significant differences in seismic inversions for
the solar temperature and hydrogen abundance profiles.
Recently, Bahcall et al.~(2004b) have also compared models constructed
using OPAL and OP opacities and they also find similar differences
in solar models due to opacities.

Although, the change in mixture from GS98 to Asp04 leads to large
changes in the inferred solar $X$ and $T$ profiles, these results do not
have much  significance because the value of $X$ near the base of the convection zone
obtained with the Asp04 mixture
is inconsistent with helioseismic estimate of $X$  in the convection zone.
The estimate of $X$ in the convection zone is barely affected by 
abundances of heavy elements (Basu \& Antia 2004).
The discrepancy in the $X$ profile is  yet another measure of the 
inconsistency between opacity, abundances and seismic constraints.

\section{Conclusions}

We find that solar envelope models that have reduced abundances of Oxygen 
and related elements do not have the correct density profile in the CZ despite
having the seismically determined CZ depth and surface $X$.
The density difference is about 15\%, which is more than 10 times
the estimated uncertainties in density.
In order to get a seismically consistent solar model it is necessary
to increase the opacities near the base of the convection zone.
The required increase in opacity is 25\% for OP tables and 27\%
for OPAL tables. The slight increase in this estimate as compared
to that by Basu \& Antia (2004) and Bahcall et al.~(2004a) is due to the
 further reduction of the 
abundance estimates for some elements. Bahcall et al.~(2005) who used
evolutionary solar models, estimate that an increase in OPAL opacity
by 11\% may be enough to get solar models in reasonable agreement
with seismic constraints. 
The difference is most likely due to the fact that the envelope Helium
abundance in these models is somewhat low (0.243) compared with
seismic abundances, and because evolutionary standard solar
models do not have  abundance profiles that agree with 
the seismically determined abundance profiles (Antia \& Chitre 1998)
in the region just below the base of the CZ.
These reasons are in addition to the further reduction
of the abundances since the work of  Bahcall et al.~(2005).
If we construct
solar envelope models with $Y=0.243$, then the required opacity
modification reduces by 5\%. 
The remaining difference is almost
certainly 
due to difference in composition profile just below the base of
CZ and further reduction in abundances of some elements.

Considering the excellent agreement between OPAL and OP opacities
it is unlikely that the error in computed opacity is of order of
10\% or larger. Thus the discrepancy between solar model with latest
abundances and seismically
inverted density profile is not likely to be due to opacities.
An independent study of solar abundances is required to
verify the recently estimated values.
One possibility is that abundance of some element has been
underestimated. To study the effect of each element separately
we have constructed models with reduction in abundance of
only one element at a time. From this study we find that the required
opacity modification is mainly controlled by abundances of O, Ne and Fe.
Thus it would be worthwhile to determine the abundances of these
elements independently to estimate any possible systematic errors in
their determination.
Of these the photospheric abundance of Ne has not been
determined spectroscopically and hence the uncertainties could be
high. Thus we speculate that the Ne abundance may be increased
to compensate for reduction in abundances of other elements.
We find that the required increase is $0.63\pm0.06$ dex for
OP opacities and $0.67\pm0.06$ dex for OPAL opacities. Thus
the estimated Ne abundance [Ne/H]$\approx8.44$ may be
compared with a value of 7.84 (Asplund et al.~2004b) and
$8.08\pm0.06$ (Grevesse \& Sauval 1998).
Of course it is unlikely that the  entire discrepancy
is due to Ne abundance, and almost certainly a part of the
discrepancy is  due to other uncertainties, including those in 
abundances of other elements. For example, if we construct models
with abundances of C, N, O, Fe increased by 0.05 dex (which is the
$1\sigma$ error estimate in their abundances) over the values
obtained by Asplund et al.~(2004b), the Ne abundance needs
to be increased by only a factor of 2.5 (0.40 dex) to get
the density within 1.5\% of the inverted values in the lower
part of the CZ. It may be noted that
we have increased the abundances of C, N, O by the same amount since these
abundances  are  correlated.
The required increase in Ne abundance is comparable to the
factor of 1.74 (0.24 dex) decrease in Ne abundance between
Grevesse \& Sauval (1998) and Asplund et al.~(2004b).

\acknowledgments

We thank the OPAL and OP groups for the  opacity tables.
This work was supported in part by NSF grants ATM 0206130 and ATM 0348837 to SB.

\clearpage

\begin{figure}
\plotone{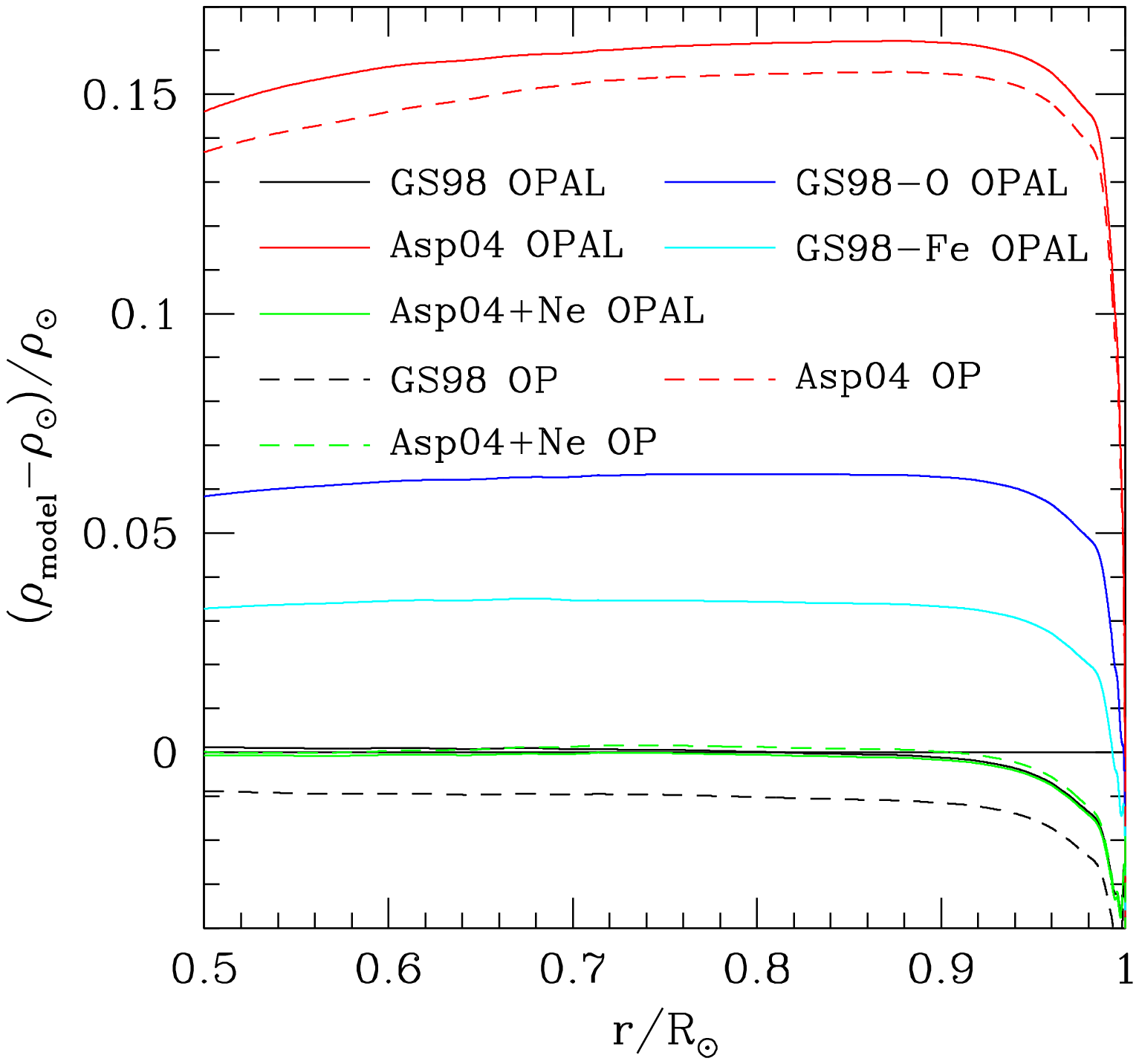}
\caption{The relative difference in density between different
solar envelope models and the Sun.
The models are constructed using different heavy element mixtures
and using either OPAL or OP opacity tables as marked in the figure.
The model Asp04+Ne is with enhanced Ne abundance in Asp04 mixture,
while GS98-O and GS98-Fe are with reduced O and Fe abundance
respectively, in GS98 mixture.
The curve for GS98 OPAL has merged with that for Asp04+Ne OPAL and
hence is not clearly visible.
\label{fig1}}
\end{figure}

\clearpage
\begin{figure}
\plotone{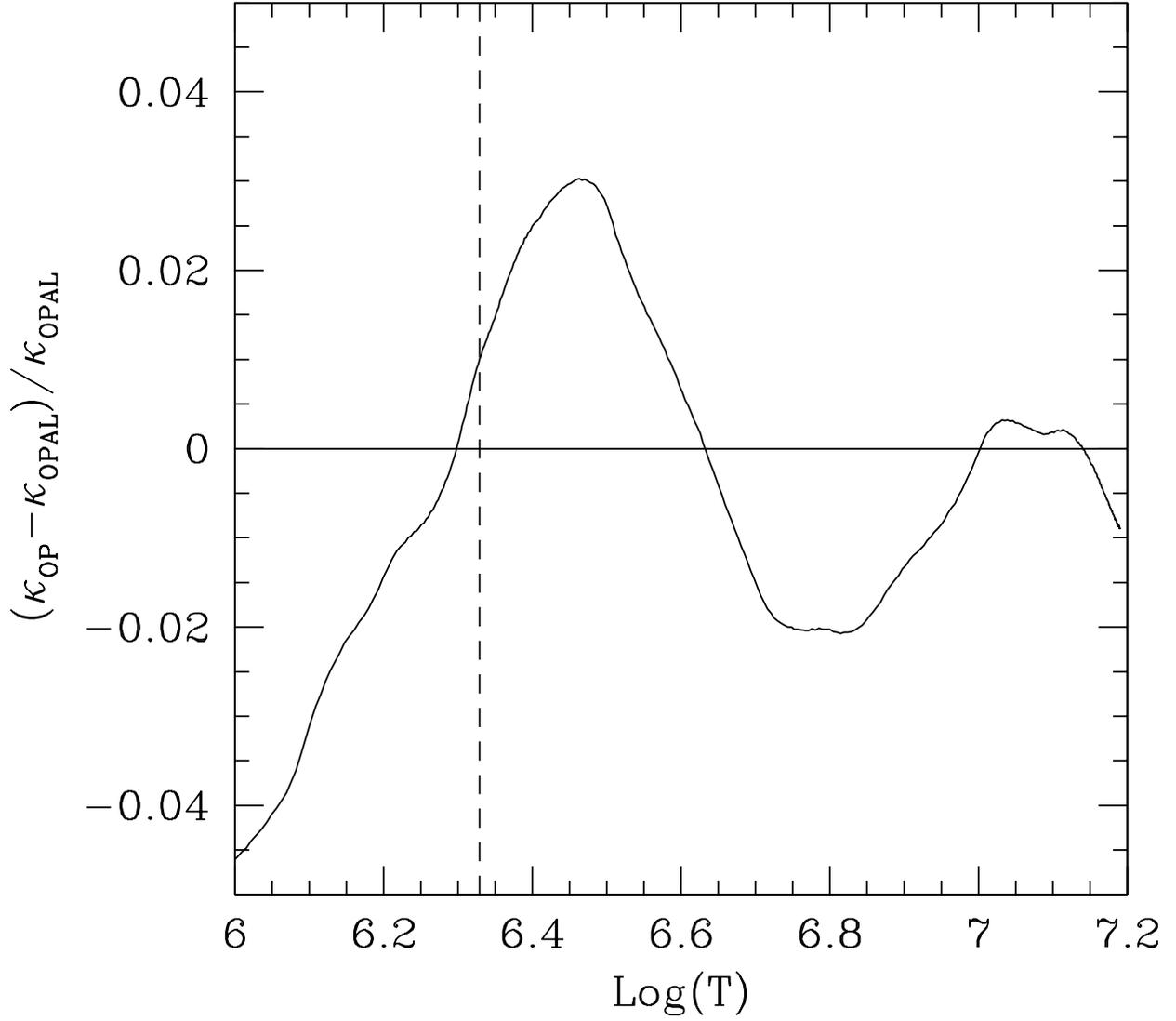}
\caption{The relative difference in opacity between OP and OPAL
tables at the same density, temperature and composition.
The dashed vertical line denotes the location of CZ base in the
solar model used.
\label{fig3}}
\end{figure}

\clearpage
\begin{figure}
\plotone{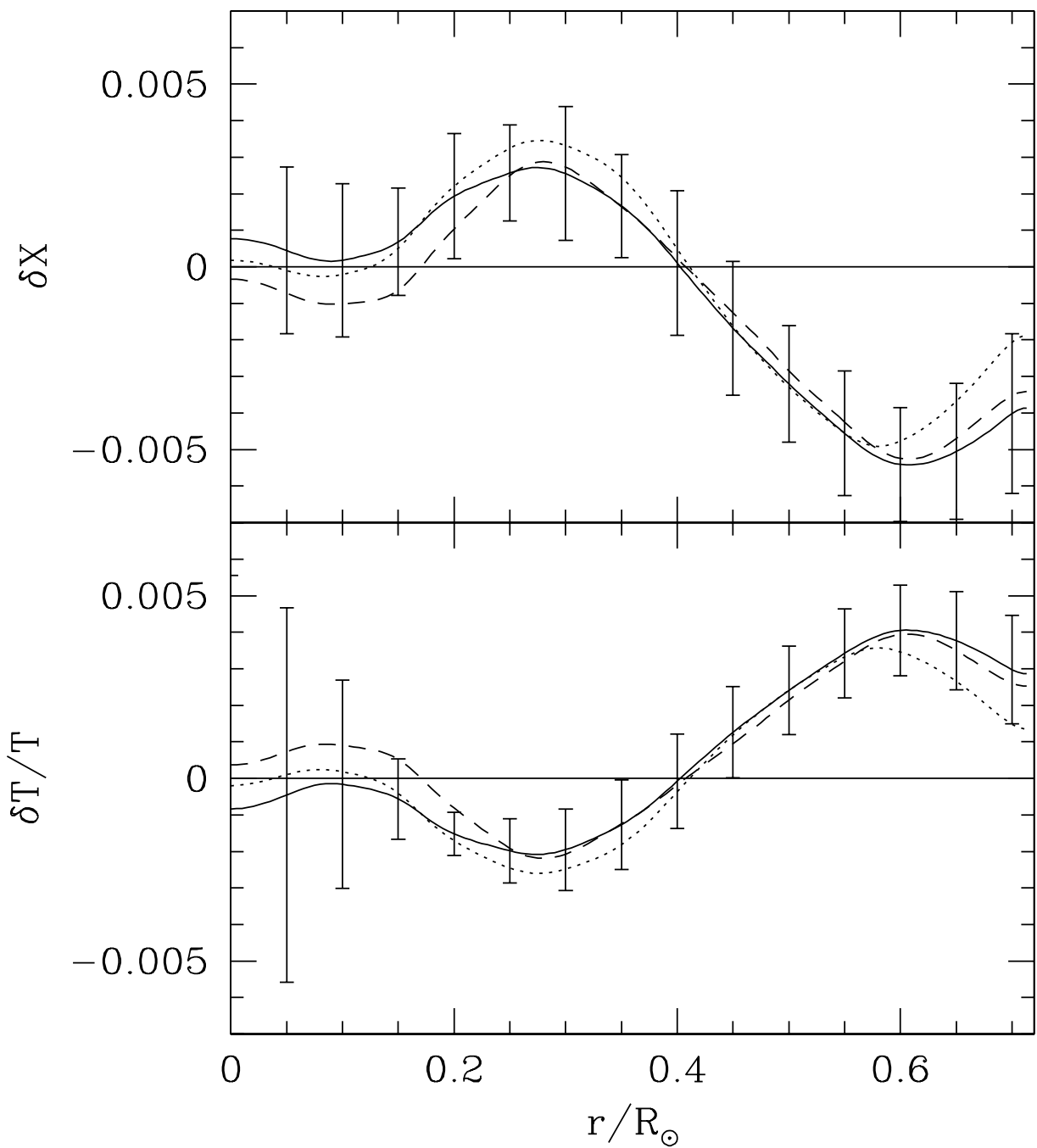}
\caption{The difference in seismically inferred
temperature and hydrogen abundance profiles between values
inferred using the OPAL and OP opacities.
The continuous lines are using the GS98 mixture, the dotted lines using
the Asp04 mixture while the dashed lines
are using Asp04 mixture with Ne abundance enhanced by 0.6 dex.
For clarity errorbars are shown on only one curve and they do not
include systematic errors.
\label{fig4}}
\end{figure}

\clearpage

\begin{table}
  \begin{center}
    \caption{Properties of different heavy element
mixtures analyzed and the required modifications in opacity
to satisfy seismic constraints. The third column gives the extent
by which the abundance of the corresponding element is reduced.
The last column gives the partial
derivative of the required opacity with respect to the abundance of
each element considered.}\vspace{1em}
    \renewcommand{\arraystretch}{1.2}
    \begin{tabular}[h]{lcccc}
      \hline
Mixture & $Z/X$ & $\delta [X/H]$ &  $\kappa/\kappa_{\rm OPAL}$ & 
$\partial\log \kappa/\partial\log X_i$\\
      \hline
GS98 & 0.0231 & & $1.000\pm0.025$ &  \\
\noalign{\smallskip}
Asp04& 0.0165 & & $1.270\pm0.030$ &  \\
\noalign{\smallskip}
GS98-C&0.0222 & 0.11 & $1.010\pm0.025$ &  $-0.04$\\
GS98-N&0.0228 & 0.12 & $1.010\pm0.025$ &  $-0.04$\\
GS98-O&0.0196 & 0.17 & $1.102\pm0.028$ &  $-0.25$\\
GS98-Ne&0.0220& 0.24 & $1.060\pm0.025$ &  $-0.11$\\
GS98-Mg&0.0228& 0.15 & $1.012\pm0.022$ &  $-0.03$\\
GS98-Si&0.0228& 0.15 & $1.010\pm0.025$ &  $-0.03$\\
GS98-S& 0.0229& 0.15 & $1.008\pm0.022$ &  $-0.02$\\
GS98-Fe&0.0226& 0.15 & $1.055\pm0.025$ &  $-0.16$\\
      \hline \\
      \end{tabular}
    \label{tab:table}
  \end{center}
\end{table}


\begin{thebibliography}{}

\bibitem[Allende Prieto et al.(2001)]{ap01}
Allende Prieto, C., Lambert, D. L., \& Asplund, M. 2001, ApJ, 556, L63

\bibitem[Allende Prieto et al.(2002)]{ap02}
Allende Prieto, C., Lambert, D. L., \& Asplund, M. 2002, ApJ, 573, L137

\bibitem[Antia \& Chitre(1998)]{ac98} Antia, H. M., \& Chitre, S. M.
1998, A\&A, 339, 239

\bibitem[Asplund et al.(2004b)]{asp04b} Asplund, M., Grevesse, N.,
Sauval, A. J., Allende Prieto, C., \& Kiselman, D. 2004a, A\&A, 417, 751

\bibitem[Asplund et al.(2004a)]{asp04} Asplund, M., Grevesse, N.,
\& Sauval, A. J. 2004b, in Cosmic abundances as records of stellar
evolution and nucleosynthesis, eds. F. N. Bash \& T. G. Barnes,
ASP Conf. Series, (in press) (astro-ph/0410214)

\bibitem[Badnell et al.(2004)]{bad04} Badnell, N. R., Bautista, M. A.,
Butler, K., Delahaye, F., Mendoza, C., Palmeri, P., Zeippen, C. J.,
\& Seaton, M. J. 2004, astro-ph/0410744

\bibitem[Bahcall \& Pinsonneault(2004)]{bp04} Bahcall, J. N., \& Pinsonneault, M. H.
2004, Phys.\ Rev.\ Lett., 92, 121301

\bibitem[Bahcall et al.(2004a)]{bsp04a} Bahcall, J. N.,
Serenelli, A. M., \& Pinsonneault, M. H. 2004, ApJ, 614, 464

\bibitem[Bahcall et al.(2004b)]{bsp04b} Bahcall, J. N.,
Serenelli, A. M., \& Basu, S. 2004, astro-ph/0412440

\bibitem[Bahcall et al.(2005)]{bsp05} Bahcall, J. N., Basu, S.,
Pinsonneault, M. H., \& Serenelli, A. M. 2005, ApJ (in press)
(astro-ph/0407060)

\bibitem[Basu \& Antia(1995)]{ba95} Basu, S., \& Antia, H. M. 1995,
\mnras, 276, 1402
 
\bibitem[Basu \& Antia(1997)]{ba97} Basu, S., \& Antia, H. M. 1997,
\mnras, 287, 189  
 
\bibitem[Basu \& Antia(2004)]{bA97} Basu, S., \& Antia, H. M. 2004,
ApJ, 606, L85
 

\bibitem[Christensen-Dalsgaard et al.(1991)]{jcd91}
Christensen-Dalsgaard, J., Gough, D. O., \& Thompson, M. J.
1991, \apj, {378}, 413


\bibitem[Gough \& Kosovichev(1988)]{dog88}
Gough, D. O., \& Kosovichev, A. G. 1988, in
Seismology of the Sun and Sun-like Stars, eds.\ V. Domingo \&
E. J. Rolfe, ESA Publ. SP-286, p.195.

\bibitem[Grevesse \& Sauval(1998)]{gre98} Grevesse, N., \& Sauval, A. J.
1998, in Solar composition and its evolution --- from core to corona,
eds., C. Fr\"ohlich, M. C. E. Huber, S. K. Solanki, \&
R. von Steiger, Kluwer, Dordrecht, p. 161

\bibitem[Iglesias \& Rogers(1996)]{opal96}
Iglesias, C. A., \& Rogers, F. J. 1996, \apj, {464}, 943

\bibitem[Melendez(2004)]{mel04}
Melendez, J. 2004, \apj, 615, 1042


\bibitem[Rogers \& Nayfonov(2002)]{opal02} Rogers, F. J., \& Nayfonov, A.
2002, \apj, 576, 1064


\bibitem[Seaton \& Badnell(2004)]{sea04} Seaton, M. J.,
\& Badnell, N. R.  2004, MNRAS, 354, 457

\bibitem[Shibahashi \& Takata(1996)]{shi96}
Shibahashi, H., \& Takata, M. 1996, PASJ, 48, 377

\bibitem[Turck-Chi\`eze et al.(2004)]{tur04}
Turck-Chi\`eze, S., Couvidat, S., Piau, L., Ferguson, J., Lambert, P.,
Ballot, J., Garcia, R. A., \& Nghiem, P. 2004, Phys.\ Rev.\ Let.,
93, 211102

\end{thebibliography}
\end{document}